\documentclass[conference]{IEEEtran}
\IEEEoverridecommandlockouts
\usepackage[T1]{fontenc}   % force legacy Times under XeTeX/tectonic: without this, TU/ptm has no bold/italic shapes and the whole document silently falls back to Latin Modern medium
\usepackage{cite}
\usepackage{amsmath,amssymb,amsfonts}
\usepackage{algorithmic}
\usepackage{graphicx}
\usepackage{textcomp}
\usepackage{xcolor}
\usepackage{optidef}
\usepackage{bm} 			% Bold math symbols inside the equation
\usepackage{mathtools} 			% For mathematical typesetting, it includes amsmath, too
\usepackage{amssymb} 			% For mathematical symbols,
\usepackage{array} 			% Extending the array and tabular environments
\usepackage{multirow} 			% Create tabular cells spanning multiple rows
\usepackage{graphicx} 			% Required to insert images
\usepackage{lscape} 			% To make defined pages' orientation landscape
\usepackage{hyperref} 			% Extensive support for hypertext (URL)
\hypersetup{hidelinks}			% No colored boxes around links (IEEE camera-ready)
\usepackage{listings} 			% Required for insertion of code
\usepackage{xcolor} 			% For custom colors to be used
\usepackage{float} 			% Required for positioning figures and table at the exact location of LaTeX code
\usepackage[open,openlevel=1]{bookmark} % It is used to have bookmarks in the PDF file created
\usepackage{comment} 			% Selectively include/exclude portions of text
\usepackage{csquotes}			% Context sensitive quotation facilities
\usepackage{enumitem} 			% To use different labeling for enumerate
\usepackage[normalem]{ulem} 		% Package for underlining
\usepackage[super]{nth}			% nth – Generate English ordinal numbers
\usepackage{gensymb}
\usepackage{enumitem}
\usepackage{tikz}			% Framework block diagram
\usetikzlibrary{arrows.meta, positioning}
\usepackage{fancyhdr}
\def\BibTeX{{\rm B\kern-.05em{\sc i\kern-.025em b}\kern-.08em
    T\kern-.1667em\lower.7ex\hbox{E}\kern-.125emX}}

\fancypagestyle{firstpage}{%
 \fancyhf{}%
 \fancyhead[L]{10th International Artificial Intelligence and Data Processing Symposium (IDAP'26), Sept 5-6, 2026, Türkiye (Malatya – İstanbul), Philippines}
  
 \fancyfoot[L]{%%%% Copyright goes here
  \normalsize{979-8-3195-2149-1/26/\$31.00 ~\copyright2026 IEEE}
 }%
}
\begin{document}

% \title{Bayesian-Optimization-Guided Sequential Convex Programming for Fuel-Optimal Collision-Free Trajectory Design on Asteroid Eros}
\title{Physics-Informed Bayesian Optimization Warm-Starts for Sequential Convex Programming in Asteroid Surface Hopping}

\author{\IEEEauthorblockN{Baran Ek\c{s}i}
\IEEEauthorblockA{\textit{Aeronautical and Astronautical Engineering} \\
\textit{Istanbul Technical University}\\
Istanbul, Türkiye \\
eksib20@itu.edu.tr}
\and
\IEEEauthorblockN{ Tufan Kumbasar} 
\IEEEauthorblockA{\textit{Artificial Intelligence and Intelligent Systems Lab.} \\
\textit{Istanbul Technical University}\\
Istanbul, Türkiye \\
kumbasart@itu.edu.tr}\\
}

\maketitle

%%%%%%%%%%%%%%%% Needs to be added for header information
% \thispagestyle{firstpage}
%%%%%%%%%%%%%%%%%%%%%%%%%%%%%%%%%%%%%

\begin{abstract}
Surface hopping is an attractive mobility mode for small-body exploration, but designing fuel-optimal hops on asteroid 433~Eros requires solving a nonconvex optimal control problem with an irregular polyhedral gravity field, thrust--mass coupling, and collision-avoidance constraints. We show that a physics-informed Bayesian Optimization (BO) warm-start---a Gaussian-process search over a single B\'{e}zier control point that requires no offline training---provides a more reliable initialization for Sequential Convex Programming (SCP) than the standard straight-line guess. The straight chord penetrates the asteroid on every inter-site transfer considered here and degrades convergence. The physics-informed BO reference feeds an SCP stage in which a log-mass change of variables convexifies the thrust--mass coupling and nearest-facet half-spaces enforce collision avoidance, and an automated Pareto time-of-flight sweep selects fuel-priority solutions without introducing bilinear terms. Applied to all 20 ordered transfers among five representative surface sites and validated under 10 random seeds, the framework tracks every trajectory to meter-level terminal accuracy in closed-loop Monte Carlo simulation, completes a five-site tour for one fifth of the propellant budget, and reduces mean $\Delta V$ by roughly one third relative to an idealized two-impulse ballistic trajectory estimate. A straight-line ablation credits the warm-start with cutting the mean SCP iteration count from 8.8 to 5.9 and removing the one convergence failure. A target-perturbation analysis further shows that the solutions vary smoothly with the landing target, with no jumps between local basins.
\end{abstract}


\section{Introduction}

Small-body exploration is moving from single-point contact toward surface mobility. The MASCOT lander relocated across Ryugu via an internal swing-arm hopping mechanism during the Hayabusa2 mission~\cite{ho2017,hayabusa2019}, and NEAR Shoemaker closed out its mission by touching down on 433~Eros~\cite{veverka2001}; neither performed a controlled, repeated transfer between chosen surface sites. That capability matters because the alternatives are poor: wheeled locomotion stalls in milligravity, and an uncontrolled ballistic hop is too imprecise to reach a specific target across kilometers of irregular terrain. Powered hopping closes this gap while requiring extra considerations about propellant consumption. 

Optimizing a single surface hop is intrinsically nonconvex due to three coupled effects: the irregular polyhedral gravity field~\cite{werner1996}, thrust dynamics coupled to propellant depletion, and continuous collision-avoidance constraints imposed by the asteroid's geometry. The importance of explicit terrain avoidance was underscored by OSIRIS-REx, which revealed Bennu's surface to be far rougher and more boulder-strewn than predicted from pre-mission observations~\cite{osiris2019}. Unlike planetary landing, hopping on a small, irregular, rotating body operates in a distinct dynamical regime~\cite{scheeres2012}; although lift-off conditions have been characterized for arbitrary shapes and rotation rates~\cite{vanwal2017}, the powered transfer must remain feasible with respect to the body's nonconvex geometry throughout the trajectory, not merely at the lift-off and landing sites.

Convex Programming (CP) provides an effective framework for tackling these nonconvexities. Lossless convexification was originally developed to convexify thrust and mass-depletion dynamics for Mars-powered descent~\cite{acikmese2007_2} and later extended to control- and pointing-constrained guidance problems~\cite{acikmese2011,acikmese2013}. Residual nonconvexities are handled by Sequential Convex Programming (SCP), which iteratively linearizes the dynamics and constraints about a reference trajectory~\cite{mao2017,szmuk2016,malyuta2022}. These methods have since been adopted for asteroid missions: fuel- and time-optimal landing~\cite{pinson2018,yang2017}, six-degree-of-freedom descent~\cite{zhang2021_6dof}, and collision-free long-range hopping~\cite{liu2021}.

The catch is that SCP only refines the reference it is handed; it does not discover a feasible corridor on its own. A straight-line reference is adequate on a quasi-spherical body, but Eros is not one: at $34.4\times11.2\times11.2$~km, the straight chord between two surface sites passes through the interior of the body on every transfer considered in this paper, so SCP starts from a deeply infeasible region and either spends extra iterations recovering or settles
 into a poor local optimum. The three-sphere keepout approximation above sidesteps this at the cost of restricting the reference away from the low-altitude corridors near the true surface. Learned references built on neural surrogates~\cite{banerjee2020,sanchez2018} are no easy fix either, requiring an offline training phase and a dataset of prior solved trajectories. What is missing
is an initialization that respects the body's actual geometry, costs almost nothing to compute online, and needs no training data at all.

This paper addresses this gap with a three-component framework applied to the complete transfer catalog among five representative sites on Eros. The contributions are:
\begin{itemize}
    \item a physics-informed Bayesian Optimization (BO) warm-start that tunes a single B\'ezier control point through an inverse-dynamics cost with no offline training phase;
    \item collision avoidance via nearest-facet half-spaces on the 1708-face polyhedron, preserving the low-altitude corridors that few-sphere bounding approximations exclude;
    \item automated time-of-flight selection: a discrete sweep whose fuel-priority solution is picked by utopia-point distance weighted by the inverse propellant mass fraction;
    \item validation at catalog scale: 20 missions across 10 random seeds, closed-loop Monte Carlo tracking to meter-level terminal accuracy, and a target-perturbation study showing that the solutions vary smoothly with the landing target, with no basin jumps.
\end{itemize}
As the warm-start is stochastic, the seed ensemble does double duty. We show that it also acts as a cheap multi-start global search, exposing local basins of the nonconvex hop problem that a single deterministic warm-start would never reveal.

The remainder of the paper is organized as follows. Section~\ref{sec:problem} formulates the problem, Section~\ref{sec:methodology} presents the proposed framework, Section~\ref{sec:results} reports the experimental results, and Section~\ref{sec:conclusion} concludes the paper.
% The remainder of the paper is organized as follows. Section~\ref{sec:problem} formulates the problem. Section~\ref{sec:methodology} details the proposed method.Section~\ref{sec:results} reports catalog performance, the warm-start ablation, seed sensitivity, and closed-loop robustness. Section~\ref{sec:conclusion} presents the conclusions.

\section{Problem Formulation}
\label{sec:problem}

This section states the problem to be solved: Section~\ref{subsec:dynamics} develops the rotating-frame dynamics and the polyhedral gravity model, Section~\ref{subsec:constraints} assembles the constraints into the nonconvex optimal
control problem, and Section~\ref{sec:scenario} defines the mission scenario and spacecraft to which it is applied.

\subsection{Dynamical Model}
\label{subsec:dynamics}
\label{subsec:gravity}

We adopt an asteroid-fixed rotating frame, eliminating time-varying endpoints; Coriolis and centrifugal terms are linear in the state and represented exactly in the convex sub-problem. The equations of motion are:
\begin{align}
    \dot{\mathbf{r}} &= \mathbf{v} \\
    \dot{\mathbf{v}} &= -2\boldsymbol{\Omega}\mathbf{v} - \boldsymbol{\Omega}^2\mathbf{r} + \mathbf{g}(\mathbf{r}) + \frac{\mathbf{T}}{m} \\
    \dot{m}          &= -\frac{\|\mathbf{T}\|}{I_{\mathrm{sp}}\,g_0}
\end{align}
where $I_{\mathrm{sp}}$ is the specific impulse, $g_0$ is standard Earth gravity, and their product is the exhaust velocity $v_{\mathrm{ex}} = I_{\mathrm{sp}}\,g_0$ (Table~\ref{tab:sc_params}); $\boldsymbol{\Omega} \in \mathbb{R}^{3\times3}$ is the skew-symmetric matrix of $\boldsymbol{\omega} = \omega\hat{\mathbf{z}}$:
\begin{equation}
    \boldsymbol{\Omega} = \begin{bmatrix} 0 & -\omega & 0 \\ \omega & 0 & 0 \\ 0 & 0 & 0 \end{bmatrix}
\end{equation}

The gravitational acceleration $\mathbf{g}(\mathbf{r}) = \nabla U(\mathbf{r})$ uses the polyhedral model of Werner and Scheeres~\cite{werner1996}, valid below the Brillouin sphere where spherical harmonics diverge, making it well suited for low-altitude hopping. The shape model is from NEAR Shoemaker data~\cite{gaskell2008,nearpds}:
\begin{equation}
    U(\mathbf{r}) = \frac{1}{2}G\rho \left[ \sum_{e} \mathbf{r}_e \cdot \mathbf{E}_e \cdot \mathbf{r}_e\,L_e - \sum_{f} \mathbf{r}_f \cdot \mathbf{F}_f \cdot \mathbf{r}_f\,w_f \right]
\end{equation}
where $\mathbf{E}_e$, $\mathbf{F}_f$ are edge and face dyads and $L_e$, $w_f$ are the logarithmic factor and solid angle. Both $\mathbf{g}(\mathbf{r})$ and its Jacobian $\partial\mathbf{g}/\partial\mathbf{r}$ are precomputed on a 3-D grid as spline interpolants; the Jacobian is used in the SCP linearization of Section~\ref{subsec:scp}.

\begin{table}[t]
\centering
\caption{Spacecraft Parameters}
\label{tab:sc_params}
\begin{tabular}{lll}
\hline
Parameter & Symbol & Value \\
\hline
Wet mass & $m_{\mathrm{wet}}$ & 250 kg \\
Dry mass & $m_{\mathrm{dry}}$ & 220 kg \\
Maximum thrust & $T_{\mathrm{max}}$ & 10 N \\
Specific impulse & $I_{\mathrm{sp}}$ & 220 s \\
Exhaust velocity & $v_{\mathrm{ex}}$ & 2.157 km/s \\
\hline
\end{tabular}
\end{table}

\subsection{Constraints and Optimal Control Problem}
\label{subsec:constraints}
\label{sec:ocp}

Fixed boundary conditions require zero velocity at departure and arrival:
\begin{align}
    \mathbf{r}(0) &= \mathbf{r}_0, \quad \mathbf{v}(0) = \mathbf{0}, \quad m(0) = m_{\mathrm{wet}} \\
    \mathbf{r}(t_f) &= \mathbf{r}_f, \quad \mathbf{v}(t_f) = \mathbf{0}, \quad m(t_f) \geq m_{\mathrm{dry}}
\end{align}
The thrust may not be directed into the surface at ignition and touchdown~\cite{acikmese2007_2,acikmese2013}:
\begin{equation}
    \hat{\mathbf{n}}_0^\top \mathbf{T}(0) \geq 0.1\,\|\mathbf{T}(0)\|, \qquad \hat{\mathbf{n}}_f^\top \mathbf{T}(t_f) \geq 0.1\,\|\mathbf{T}(t_f)\|
\end{equation}
where the coefficient 0.1 permits thrust up to $\approx 84^\circ$ from the local surface normal, which is a weak pointing condition that forbids only surface-directed thrust.

A glide-slope cone ($\alpha = 60^\circ$) prevents surface collision during ascent and descent~\cite{acikmese2007_2}:
\begin{equation}
    \|\mathbf{r}(t) - \mathbf{r}_s\|\cos\alpha \leq \hat{\mathbf{n}}_s^\top(\mathbf{r}(t) - \mathbf{r}_s), \quad t \in \mathcal{T}_{\mathrm{as}}
\end{equation}
During cruise the trajectory must clear the surface by $d_{\mathrm{safe}} = 50$~m:
\begin{equation}
    \hat{\mathbf{n}}_k^\top\!\left(\mathbf{r}(t) - \mathbf{c}_k\right) \geq d_{\mathrm{safe}}, \quad t \in \mathcal{T}_{\mathrm{cr}}
\end{equation}
where $\hat{\mathbf{n}}_k$ and $\mathbf{c}_k$ are the normal and centroid of the nearest surface facet. Minimizing fuel for fixed $t_f$ yields the optimal control problem (OCP):
\begin{equation}
\begin{aligned}
    \min_{\mathbf{T}(t)} \quad & -m(t_f) \\
    \text{s.t.} \quad & \dot{\mathbf{x}} = f(\mathbf{x}, \mathbf{T}), \quad t \in [0, t_f] \\
    & \mathbf{r}(0) = \mathbf{r}_0,\; \mathbf{r}(t_f) = \mathbf{r}_f,\; \mathbf{v}(0) = \mathbf{v}(t_f) = \mathbf{0} \\
    & m(0) = m_{\mathrm{wet}},\; m(t_f) \geq m_{\mathrm{dry}},\; \|\mathbf{T}\| \leq T_{\mathrm{max}} \\
    & \hat{\mathbf{n}}_0^\top \mathbf{T}(0) \geq 0.1\|\mathbf{T}(0)\|,\; \hat{\mathbf{n}}_f^\top \mathbf{T}(t_f) \geq 0.1\|\mathbf{T}(t_f)\| \\
    & \|\mathbf{r}(t) - \mathbf{r}_s\|\cos\alpha \leq \hat{\mathbf{n}}_s^\top(\mathbf{r}(t) - \mathbf{r}_s), \quad t \in \mathcal{T}_{\mathrm{as}} \\
    & \hat{\mathbf{n}}_k^\top(\mathbf{r}(t) - \mathbf{c}_k) \geq d_{\mathrm{safe}}, \quad t \in \mathcal{T}_{\mathrm{cr}}
\end{aligned}
\end{equation}
Here, $\mathcal{T}_{\mathrm{as}}$ and $\mathcal{T}_{\mathrm{cr}}$ denote ascent/descent and cruise intervals; the transition occurs at 800~m from each site, a radius that bounds the terminal maneuvering region and hands the remainder of the transfer to the clearance constraint. This problem is nonconvex due to its dynamics and constraints. Because Section~\ref{subsec:scp} solves this nonconvex problem by successive linearization, the resulting trajectories are locally rather than globally optimal. The phrase "fuel-optimal" throughout this paper denotes a converged minimizer of the objective above, not a certified global minimum, and Section~\ref{subsec:seed_dep} shows this basin-dependence directly.

\subsection{Simulation Scenario}\label{sec:scenario}
\label{sec:mission}
\label{sec:spacecraft}

Tables~\ref{tab:sc_params}--\ref{tab:eros_params} summarize the spacecraft and 433~Eros parameters. Eros properties are taken from the NEAR radio science and shape solutions~\cite{yeomans2000,miller2002}, and the 856-vertex, 1708-face polyhedral model from ~\cite{nearpds}.

\begin{table}[b]
\centering
\caption{433 Eros Physical Properties}
\label{tab:eros_params}
\begin{tabular}{lll}
\hline
Parameter & Symbol & Value \\
\hline
Rotation period & $T_{\mathrm{Eros}}$ & 5.270 h \\
Rotation rate & $\omega$ & $3.312 \times 10^{-4}$ rad/s \\
Gravitational parameter & $\mu$ & $4.463 \times 10^{-4}$ km$^3$/s$^2$ \\
Bulk density & $\rho$ & 2670 kg/m$^3$ \\
Mean radius & $\bar{r}$ & 8.43 km \\
Dimensions & --- & $34.4 \times 11.2 \times 11.2$ km \\
\hline
\end{tabular}
\\[1.1em]
\caption{Representative Surface Sites}
\label{tab:sites}
\begin{tabular}{lllll}
\hline
Site & Description & $x$ (km) & $y$ (km) & $z$ (km) \\
\hline
P1 & Saddle region       & $-0.116$            & $-3.076$           & $\phantom{-}0.942$ \\
P2 & $-x$ long-axis tip  & $-17.578$           & $-1.471$           & $-0.551$ \\
P3 & $+x$ long-axis tip  & $\phantom{-}15.032$ & $-3.657$           & $\phantom{-}0.466$ \\
P4 & $+z$ apex           & $\phantom{-}8.325$  & $-2.547$           & $\phantom{-}5.859$ \\
P5 & $+y$ ridge          & $-2.354$            & $\phantom{-}8.431$ & $-0.550$ \\
\hline
\end{tabular}
\end{table}

\begin{figure}[!t]
    \centering
    \includegraphics[width=0.45\textwidth]{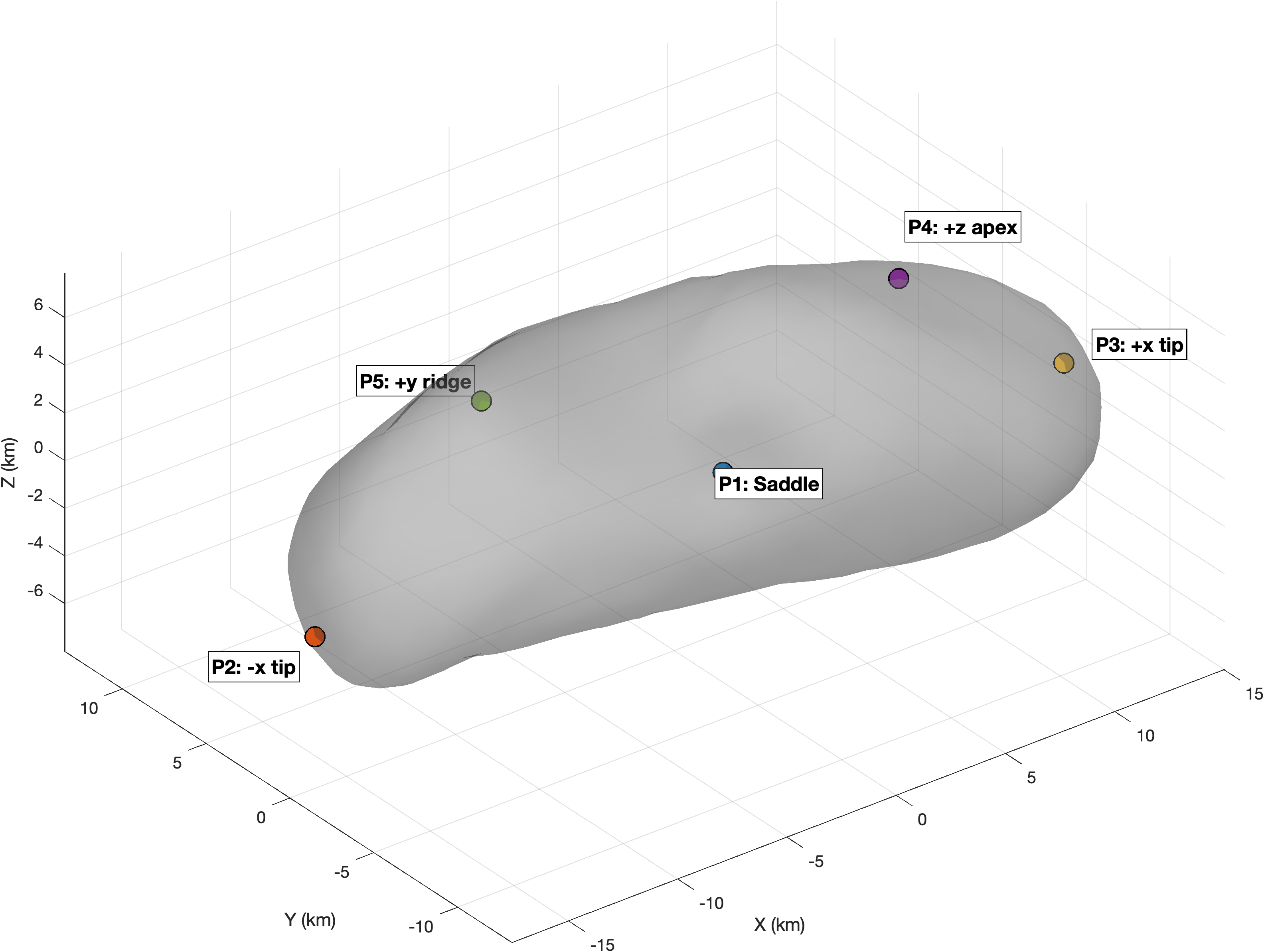}
    \caption{Representative surface sites on 433 Eros. P1--P5 are marked on the polyhedral shape model.}
    \label{fig:sites}
\end{figure}

Five geometrically distinct surface sites (Table~\ref{tab:sites}, Fig.~\ref{fig:sites}) are selected to span Eros's irregular morphology---the long-axis tips (P2, P3), the $+z$ apex (P4), the central saddle (P1), and the $+y$ ridge (P5). All $5 \times 4 = 20$ ordered departure--arrival pairs are considered.

\section{Methodology}
\label{sec:methodology}

\begin{figure}[!t]
\centering
\begin{tikzpicture}[
    node distance=3.5mm,
    box/.style={draw, rounded corners=1pt, align=center, font=\scriptsize,
                minimum width=0.52\linewidth, inner sep=3pt},
    arrlbl/.style={font=\tiny\itshape, midway, right=1.5mm},
    >=Stealth]
  \node[box] (bo)  {\textbf{Physics-informed Bayesian warm-start} (Sec.~\ref{sec:bo})\\ near-feasible B\'ezier reference};
  \node[box, below=of bo] (scp) {\textbf{Convexified SCP} (Sec.~\ref{subsec:scp})\\ converged trajectory per $t_f \in \mathcal{T}$};
  \node[box, below=of scp] (par) {\textbf{Pareto $t_f$ selection} (Sec.~\ref{sec:pareto})\\ utopia-point, fuel-priority $t_f^*$};
  \node[box, below=of par] (val) {\textbf{Closed-loop validation} (Sec.~\ref{sec:robustness})\\ 500-run Monte Carlo + target perturbation};
  \draw[->] (bo)  -- (scp) node[arrlbl] {initial reference};
  \draw[->] (scp) -- (par) node[arrlbl] {$(t_{f,k},\,m_{f,k})$ Pareto front};
  \draw[->] (par) -- (val) node[arrlbl] {nominal trajectory};
\end{tikzpicture}
\caption{The proposed framework. A physics-informed BO warm-start supplies a near-feasible initial reference, SCP solves the convexified OCP over a time-of-flight grid, a utopia-point criterion automates the time--fuel trade-off, and every solution is validated in closed loop.}
\label{fig:framework}
\end{figure}
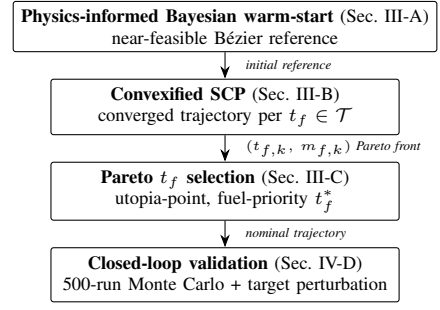

Eros's elongated, highly irregular geometry motivates the framework in Fig.~\ref{fig:framework}. A physics-informed BO warm-start first generates a near-feasible Bézier reference (Section~\ref{sec:bo}), which initializes an SCP solver for the nonconvex OCP (Section~\ref{subsec:scp}). To preserve convexity, the final time is optimized by sweeping a discrete set of $t_f$ values rather than treating it as a decision variable; the resulting Pareto solutions are ranked using a fuel-priority utopia-point criterion (Section~\ref{sec:pareto}).
% Eros's elongated geometry drives the design of the framework shown in Fig.~\ref{fig:framework}. A physics-informed BO warm-start first constructs a near-feasible B\'ezier reference (Section~\ref{sec:bo}), and the nonconvex OCP is then solved by SCP around this reference (Section~\ref{subsec:scp}). Since a free final time would introduce bilinear terms and destroy convexity, $t_f$ is swept over a discrete grid, and a fuel-priority solution is selected by a utopia-point Pareto criterion (Section~\ref{sec:pareto}).

\subsection{Physics-informed Bayesian Warm-Start}
\label{sec:bo}

Straight-line initializations geometrically penetrate Eros on every inter-site transfer of the catalog, degrading the convergence of the SCP stage that follows. The physics-informed BO warm-start instead generates a near-feasible initial reference that minimizes surface penetration. It is physics-informed in the sense that its cost is computed from the hop dynamics themselves rather than from a learned model, so unlike learning-based warm-starts~\cite{banerjee2020,sanchez2018}, it requires no offline training data. The trajectory is parameterized as a quadratic B\'ezier curve with fixed endpoints $\mathbf{P}_0 = \mathbf{r}_0$, $\mathbf{P}_2 = \mathbf{r}_f$ and a free control point $\mathbf{P}_1 \in [-15,\,15]^3$~km:
\begin{equation}
    \mathbf{r}(\tau) = (1-\tau)^2\mathbf{P}_0 + 2(1-\tau)\tau\,\mathbf{P}_1 + \tau^2\mathbf{P}_2, \quad \tau \in [0,1]
\end{equation}
Optimizing only $\mathbf{P}_1$ reduces the warm-start to a three-parameter search; the curve responds to $\mathbf{P}_1$ with gain $2(1-\tau)\tau \leq 1/2$, so the box can displace the mid-arc by up to $7.5$~km in any direction, comparable to Eros's own extent (Table~\ref{tab:eros_params}). The physics-informed BO cost evaluates each candidate via inverse dynamics:
\begin{equation}
    \mathbf{u}_{\mathrm{req},k} = \ddot{\mathbf{r}}_k + 2\boldsymbol{\Omega}\dot{\mathbf{r}}_k + \boldsymbol{\Omega}^2\mathbf{r}_k - \mathbf{g}(\mathbf{r}_k)
\end{equation}
\begin{multline}
    J_{\mathrm{BO}} = w_1\!\sum_{k\in\mathcal{S}}\|\mathbf{u}_{\mathrm{req},k}\|^2 + w_2\!\sum_{k=0}^{N}c(\phi_k) \\
    + w_3\!\sum_{k\in\mathcal{S}}\max\bigl(0,\|\mathbf{u}_{\mathrm{req},k}\|-u_{\mathrm{max}}\bigr)^2
\end{multline}
where $\mathcal{S}$ samples every fifth node, $\phi_k$ is the signed surface clearance of node $k$, and $u_{\mathrm{max}} = T_{\mathrm{max}}/m_{\mathrm{wet}}$ is the maximum available thrust acceleration. The weights $(w_1,w_2,w_3) = (0.1,\,100,\,1000)$ are ordered so that any collision penalty ($w_2$) or thrust-infeasibility penalty ($w_3$) dominates the control-effort term ($w_1$); their exact values are not critical, as the penalty terms act as soft barriers. The clearance penalty is:
\begin{equation}
    c(\phi_k) = \begin{cases} 10^6 + e^{20|\phi_k|} & \phi_k < 0 \\ e^{20(\phi_{\mathrm{buf}}-\phi_k)} & 0 \leq \phi_k < \phi_{\mathrm{buf}} \\ 0 & \phi_k \geq \phi_{\mathrm{buf}} \end{cases}
\end{equation}
with $\phi_{\mathrm{buf}} = 0.5$~km. Between colliding and clear candidates this cost spans several orders of magnitude; the surrogate need only rank the colliding region as unpromising, which the 16-of-20 clearing rate reported below confirms.

MATLAB's \texttt{bayesopt} minimizes $J_{\mathrm{BO}}$ by BO with a Gaussian process surrogate and the EI+ acquisition function ($\epsilon$-greedy expected improvement~\cite{bull2011}) over 50 evaluations. The warm-start completes in roughly 10~s---about one SCP iteration---and is repaid several times over by the iterations it saves.

A seed-0 characterization at this budget shows the resulting reference clears the body on 16 of the 20 transfers and reduces the worst-case surface penetration of the reference from $4.5$~km (straight chord) to at most $0.92$~km. On the remaining transfers, no clearing quadratic B\'ezier was found, and the residual infeasibility is absorbed by the slack variables of the first SCP iterations. The optimal $\mathbf{P}_1^*$ defines the position reference along the B\'ezier arc. Since BO is stochastic, each mission pair is solved under 10 independent seeds, numbered 0 through 9, each initializing the random number generator prior to the warm-start, yielding $20 \times 10 = 200$ reproducible optimization runs.

\subsection{Sequential Convex Programming}
\label{subsec:scp}

Successive convexification---the SCP variant adopted here~\cite{mao2017,szmuk2016,malyuta2022}---handles each nonconvexity source of the OCP by iterative linearization around a position and log-mass reference $(\bar{\mathbf{r}}_k, \bar{z}_k)$, yielding a convex second-order cone program (SOCP) per iteration. The position reference is supplied by the warm-start of Section~\ref{sec:bo}; the log-mass reference is initialized by linear interpolation from $\ln m_{\mathrm{wet}}$ to $\ln m_{\mathrm{dry}}$.

\textbf{Thrust--mass convexification.} Following the change of variables of~\cite{acikmese2007_2,acikmese2011}, the control is the mass-normalized thrust acceleration $\mathbf{u}_k = \mathbf{T}_k/m_k$ and the mass is replaced by its logarithm $z = \ln m$. A slack $\sigma_k \geq 0$ bounds the control magnitude and renders depletion linear ($\dot{z} = -\sigma_k/v_{\mathrm{ex}}$); since the minimum thrust is zero, the relaxation $\|\mathbf{u}_k\| \leq \sigma_k$ is exact, and only the mass-dependent ceiling requires linearization about $\bar{z}_k$:
\begin{equation}
    \|\mathbf{u}_k\| \leq \sigma_k, \qquad \sigma_k \leq T_{\mathrm{max}}\,e^{-\bar{z}_k}\bigl(1 - (z_k - \bar{z}_k)\bigr)
\end{equation}

\textbf{Discretization and linearization.} The trajectory is discretized into $N = 300$ intervals ($\Delta t = t_f/N$). Gravity is linearized via first-order Taylor expansion at interval midpoints, with Jacobian $\mathbf{J}_k = \partial\mathbf{g}/\partial\mathbf{r}\big|_{\bar{\mathbf{r}}_{mid,k}}$ from the precomputed interpolant of Section~\ref{subsec:gravity}. State propagation uses the trapezoidal rule with virtual control slack $\boldsymbol{\nu}_k \in \mathbb{R}^6$:
\begin{equation}
    \begin{bmatrix}\mathbf{r}_{k+1}\\\mathbf{v}_{k+1}\end{bmatrix} = \begin{bmatrix}\mathbf{r}_k\\\mathbf{v}_k\end{bmatrix} + \frac{\Delta t}{2}(\mathbf{f}_k + \mathbf{f}_{k+1}) + \boldsymbol{\nu}_k, \qquad z_{k+1} = z_k - \frac{\Delta t}{v_{\mathrm{ex}}}\sigma_k
\end{equation}
where $\mathbf{f}_k$ stacks the velocity and the acceleration (Coriolis, centrifugal, linearized gravity, and control) at node $k$.

\textbf{Collision avoidance, trust region, and convergence.} During cruise, the nearest facet of the full 1708-face polyhedron defines a half-space, a feasible corridor that follows the true surface, whereas the few-sphere bounding approximations of prior hopping work~\cite{liu2021} exclude the near-surface region where several of the fuel-optimal solutions of Section~\ref{sec:results} ride the clearance limit:
\begin{equation}
    \hat{\mathbf{n}}_k^\top(\mathbf{r}_k - \mathbf{c}_k) \geq d_{\mathrm{safe}} - s_k, \quad s_k \geq 0
\end{equation}
Position is confined to a trust region $\|\mathbf{r}_k - \bar{\mathbf{r}}_k\|_2 \leq \delta$, initialized at 20~km for the first iteration and thereafter adapted within $[0.5,\,15]$~km: contracted by $0.7\times$ when the physics infeasibility does not decrease, expanded by $1.2\times$ otherwise. Convergence requires the node-averaged position change $\|\mathbf{r} - \bar{\mathbf{r}}\|_F/N$, the largest virtual-control magnitude $\max_k \|\boldsymbol{\nu}_k\|$, and the largest safety slack $\max_k s_k$ to fall below $10^{-2}$, $10^{-3}$, and $10^{-3}$~km, respectively. The objective is:
\begin{equation}
    \min\;-z_{N+1} + W_v\!\sum_{k}\|\boldsymbol{\nu}_k\|_1 + W_p\!\sum_{k}s_k
\end{equation}
where $W_v = 10^4$ and $W_p = 10^5$, solved by CVX~\cite{cvx}, with a 25-iteration limit per flight time.

\begin{figure}[b]
    \centering
    \includegraphics[width=0.45\textwidth]{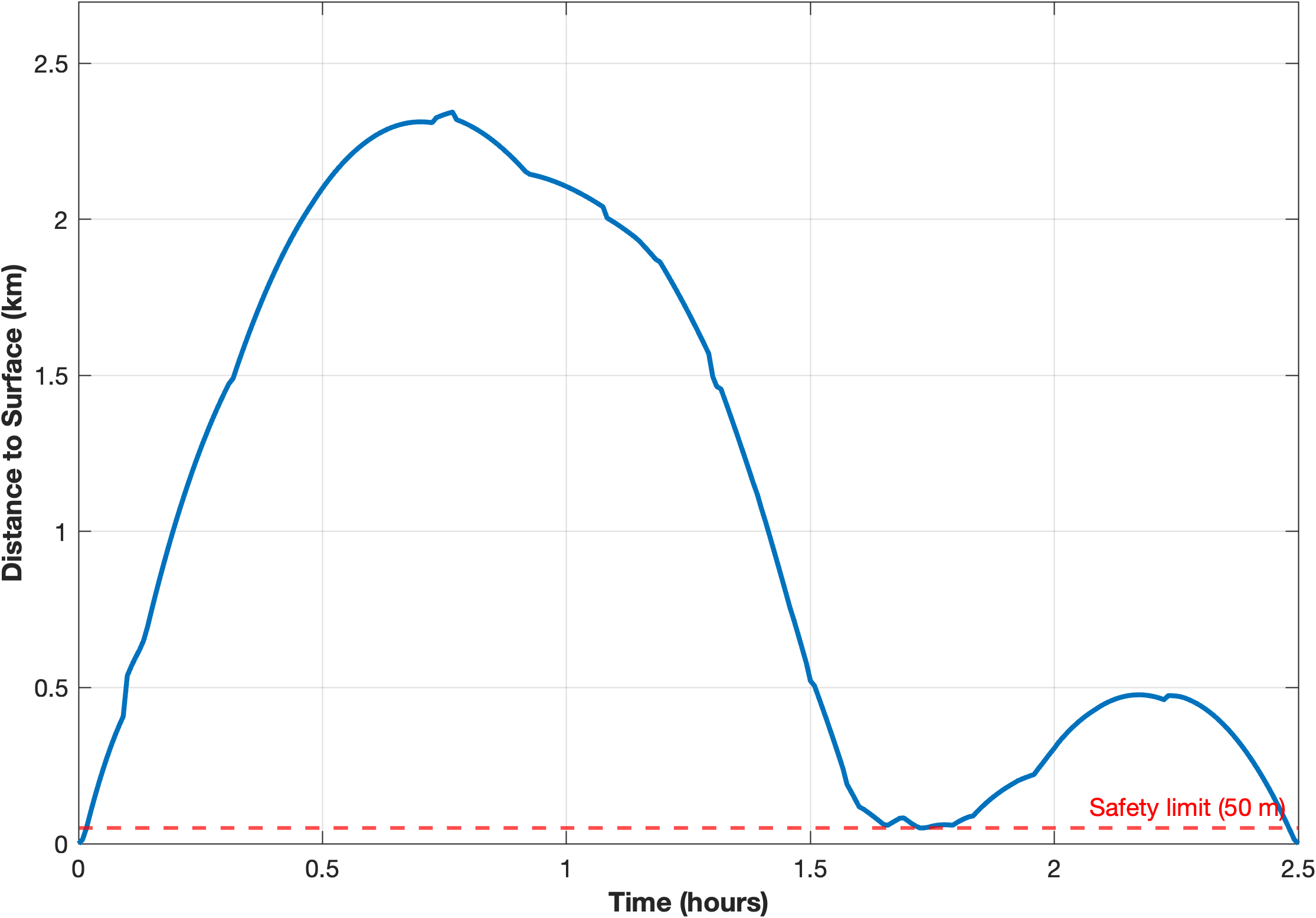}
    \caption{The $d_{\mathrm{safe}} = 50$~m clearance constraint is active: surface
    clearance profile for M13 (P5$\to$P2, $d = 18.2$~km, $t_f = 2.5$~h),
    the mission with the largest $\Delta V$ reduction in the catalog,
    which reaches the safety limit near $t \approx 1.7$~h while following
    a low-altitude corridor.}
    \label{fig:alt_m04}
\end{figure}

\begin{figure}[b]
    \centering
    \includegraphics[width=0.45\textwidth]{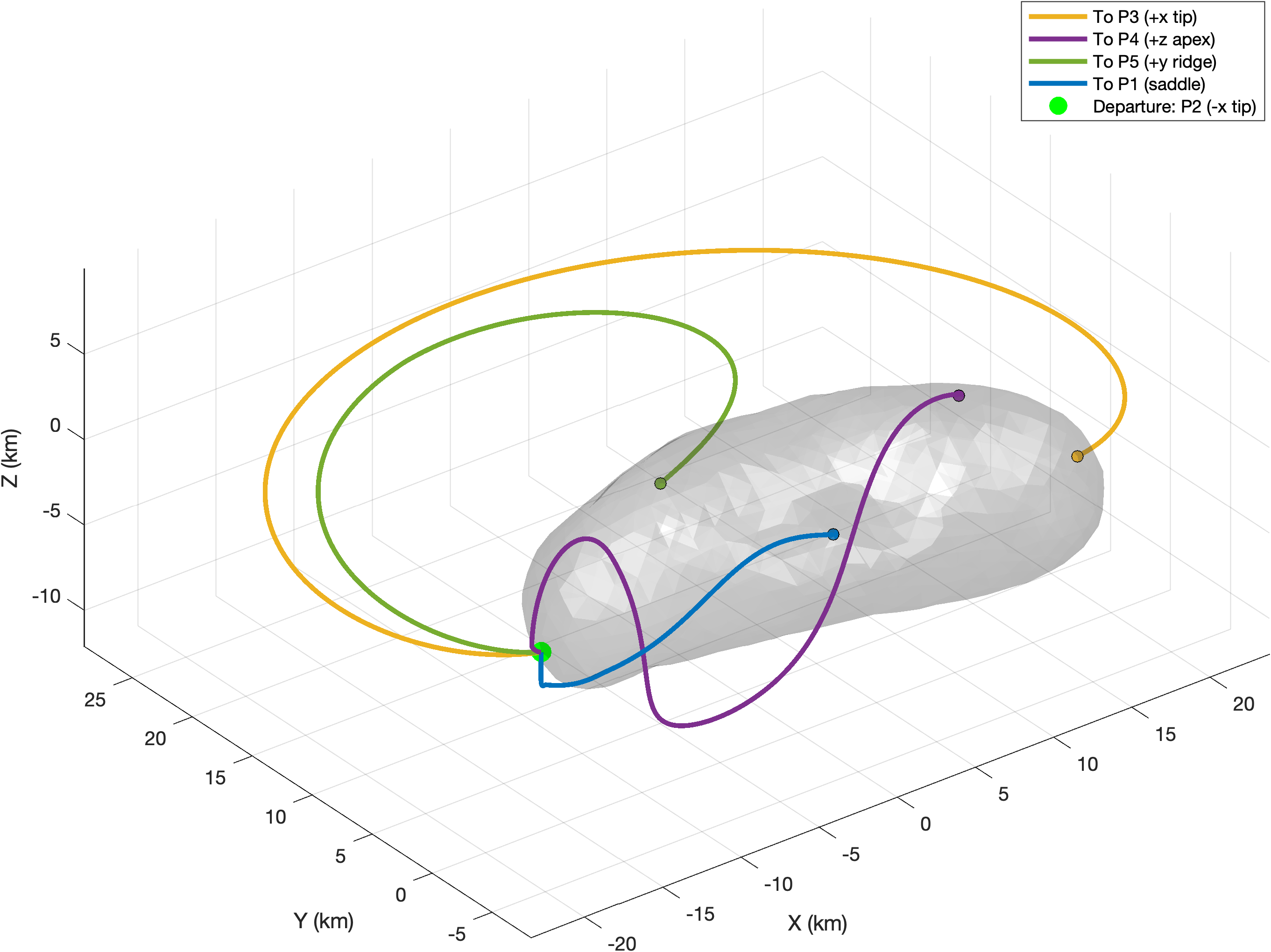}
    \caption{Fuel-optimal transfers wrap around Eros's irregular shape rather than taking direct arcs: the four missions departing P2 ($-x$ long-axis tip), to P3 (orange), P4 (purple), P5 (green), and P1 (blue); the sphere marks the departure site. All trajectories are from seed~0.}
    \label{fig:trajs_p2}
\end{figure}

\subsection{Pareto Time-of-Flight Selection}
\label{sec:pareto}

Including $t_f$ as a decision variable introduces bilinear terms $t_f\mathbf{u}$, destroying convexity. Instead, SCP is solved for each fixed $t_f$ on the grid $\mathcal{T} = \{1.5,\,1.75,\,2.0,\,2.25,\,2.5,\,3.0,\,3.5,\,4.0,\,4.5\}$~h, each solve warm-started from the previous converged solution. The resulting $(t_{f,k},\,m_{f,k})$ pairs form a discrete trade-off front; sweep points dominated in both objectives cannot minimize the utopia distance of \eqref{eq:utopia} and are therefore never selected. Objectives are normalized to $[0,1]$ and each candidate is ranked by weighted utopia-point distance:
\begin{equation}
    d_k = \sqrt{\bar{t}_k^2 + w_m(1-\bar{m}_k)^2}, \quad w_m = \frac{m_{\mathrm{wet}}}{m_{\mathrm{wet}}-m_{\mathrm{dry}}}
    \label{eq:utopia}
\end{equation}
where $\mathrm{PMF}=(m_{\mathrm{wet}}-m_{\mathrm{dry}})/m_{\mathrm{wet}}$ is the Propellant Mass Fraction (PMF). Weighting the fuel axis by the inverse PMF ties fuel priority directly to propellant shortage. The perpendicular-distance knee was rejected: for Eros it places the knee at short flight times, selecting high-fuel solutions. The utopia-point criterion then selects the fuel-priority candidate:
\begin{equation}
    t_f^* = \operatorname*{arg\,min}_{t_{f,k}\in\mathcal{T}} d_k
\end{equation}

\section{Results and Discussion}
\label{sec:results}

Results are reported in four parts. Section~\ref{subsec:nominal} presents the 20-mission catalog. Section~\ref{subsec:ablation} isolates the contribution of the warm-start by ablation, Section~\ref{subsec:seed_dep} quantifies sensitivity to its stochasticity across the 10 seeds, and Section~\ref{sec:robustness} validates the selected trajectories in closed loop.

\subsection{Catalog Performance}
\label{subsec:nominal}

\begin{table*}[t]
\centering
\caption{Per-Mission Results Across 10 Seeds (200 Runs)}
\label{tab:results}
\begin{tabular}{clccccccc}
\hline
ID & Pair & $d$ (km) & $t_f$ (h) & Fuel (kg) & $\Delta V_{\mathrm{SCP}}$ (m/s) & $\Delta V_{\mathrm{bal}}$ (m/s) & $\Delta$ (\%) & Iters \\
\hline
M01 & P1$\to$P2 & 17.6 & 4.00          & 1.434           & 12.4 & 21.0 & $-41.0$ & 2.0 \\
M02 & P1$\to$P3 & 15.2 & $1.88\pm0.24$ & $1.453\pm0.328$ & 12.6 & 19.5 & $-35.5$ & 4.8 \\
M03 & P1$\to$P4 &  9.8 & 1.50          & $1.779\pm0.001$ & 15.4 & 15.7 & $-1.7$  & 6.7 \\
M04 & P1$\to$P5 & 11.8 & 4.50          & 1.782           & 15.4 & 17.2 & $-10.4$ & 7.0 \\
M05 & P2$\to$P3 & 32.7 & 4.00          & 1.486           & 12.9 & 28.7 & $-55.1$ & 2.0 \\
M06 & P2$\to$P4 & 26.7 & $3.55\pm0.60$ & $1.602\pm0.089$ & 13.9 & 25.9 & $-46.4$ & 11.6 \\
M07 & P2$\to$P5 & 18.2 & 4.00          & 1.284           & 11.1 & 21.4 & $-48.0$ & 8.0 \\
M08 & P3$\to$P4 &  8.7 & 1.50          & $1.475\pm0.001$ & 12.8 & 14.8 & $-13.5$ & 11.3 \\
M09 & P3$\to$P5 & 21.2 & $2.20\pm0.11$ & $1.072\pm0.009$ &  9.3 & 23.1 & $-59.8$ & 3.0 \\
M10 & P4$\to$P5 & 16.6 & 2.50          & 1.668           & 14.5 & 20.4 & $-29.3$ & 4.0 \\
M11 & P5$\to$P4 & 16.6 & 3.00          & 1.624           & 14.1 & 20.4 & $-31.1$ & 3.0 \\
M12 & P5$\to$P3 & 21.2 & 3.50          & 1.432           & 12.4 & 23.1 & $-46.3$ & 2.0 \\
M13 & P5$\to$P2 & 18.2 & 2.50          & $0.964\pm0.002$ &  8.3 & 21.4 & $-61.0$ & 4.0 \\
M14 & P5$\to$P1 & 11.8 & 1.50          & $1.831\pm0.036$ & 15.9 & 17.2 & $-7.9$  & 4.8 \\
M15 & P4$\to$P3 &  8.7 & 2.50          & 1.416           & 12.3 & 14.8 & $-17.0$ & 14.0 \\
M16 & P4$\to$P2 & 26.7 & 3.50          & $1.653\pm0.001$ & 14.3 & 25.9 & $-44.7$ & 4.0 \\
M17 & P4$\to$P1 &  9.8 & 1.50          & 1.739           & 15.1 & 15.7 & $-3.9$  & 3.0 \\
M18 & P3$\to$P2 & 32.7 & 3.50          & 1.413           & 12.2 & 28.7 & $-57.3$ & 2.0 \\
M19 & P3$\to$P1 & 15.2 & $4.25\pm0.26$ & $1.533\pm0.043$ & 13.3 & 19.5 & $-32.0$ & 4.0 \\
M20 & P2$\to$P1 & 17.6 & $1.80\pm0.16$ & $1.340\pm0.005$ & 11.6 & 21.0 & $-44.9$ & 3.8 \\
\hline
\end{tabular}
\end{table*}

The framework solves the full 20-mission catalog on every seed, and every mission is cheaper than its ballistic reference. Table~\ref{tab:results} reports per-mission results across 10 seeds (200 total runs): $t_f$ and fuel as mean~$\pm$~std, with bare values indicating zero inter-seed spread; the $\Delta V$, $\Delta$, and iteration columns report seed means. Mean fuel consumption per hop is $1.50$~kg and the full catalog sums to $29.98 \pm 0.33$~kg across seeds. Since the catalog enumerates all $20$ ordered site pairs, this total characterizes vehicle capability rather than costing a mission: an actual mission flies a subset. A representative five-site tour P1$\to$P2$\to$P3$\to$P4$\to$P5 (missions M01, M05, M08, M10) costs $6.06$~kg, $20.2\%$ of the propellant budget, identical on all 10 seeds. Moreover, chained hops depart lighter than the $m_{\mathrm{wet}}$ assumed for each catalog entry, making these per-hop figures conservative.

The baseline $\Delta V_{\mathrm{bal}}$ is a flat-body two-impulse ballistic estimate, $\Delta V = 2\sqrt{g_s d}$ for chord length $d$ at the optimal $45^\circ$ launch angle with constant mean surface gravity $g_s = \mu/\bar{r}^2$---not a flyable transfer on a rotating irregular body, but an intuitive impulsive-$\Delta V$ scale (the $\Delta$ column of Table~\ref{tab:results}; negative favors SCP). Against it, SCP achieves 13.0~m/s mean $\Delta V$ versus 20.8~m/s, a 34\% mean per-mission reduction, cheaper on all 20 missions. The saving is largest on the longer transfers, reaching $55$--$61\%$ (M05, M09, M13, M18), where SCP exploits the irregular gravity field through low-altitude corridors; on the short hops between the saddle and the apex or ridge (M03, M04, M14, M17) the powered cost approaches the impulsive estimate; these short, terrain-dominated transfers are precisely where the flat-body assumption underestimates the true cost.

The SCP solver converges in a median of 4 iterations (mean 5.2, maximum 19); the short, terrain-constrained P3$\leftrightarrow$P4 pair (M08, M15) and the long P2$\to$P4 transfer (M06) require the most, averaging 11--14. Of the 1800 sweep-point solves, only $1.8\%$ hit the 25-iteration cap, concentrated at two systematically hard mission/flight-time pairs (M08 at 2.0~h, M10 at 4.5~h) that the Pareto criterion never selects. Per-mission wall-clock time averages 5.8~min for the full nine-point sweep; the 20-mission catalog completes in approximately $2$~h on an Apple~M2 computer (8~GB RAM, MATLAB R2024b).

\begin{figure}[t]
    \centering
    \includegraphics[width=0.45\textwidth]{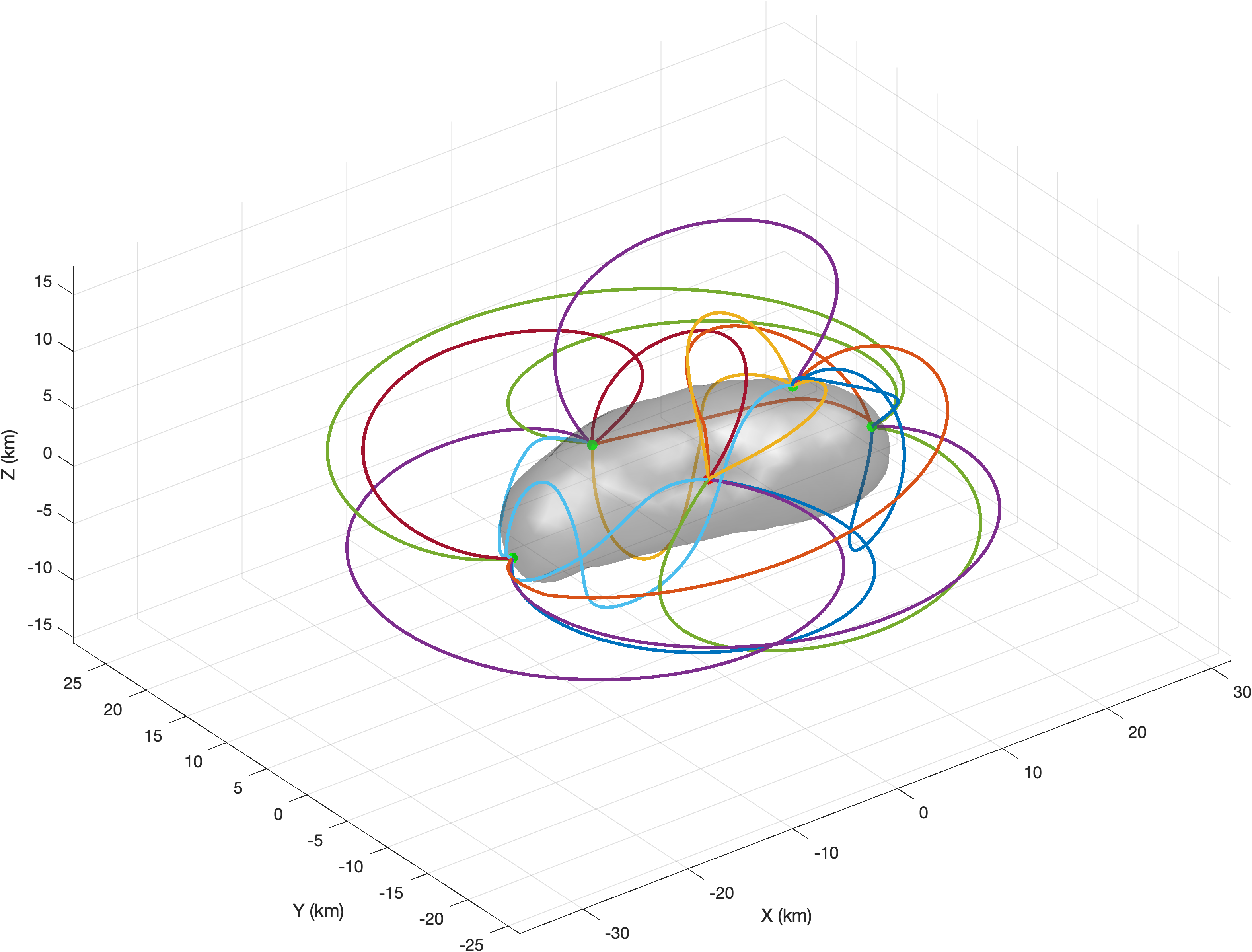}
    \caption{The complete catalog: all 20 fuel-optimal hopping trajectories
    on 433~Eros (seed~0). Colors separate the transfers visually and repeat
    across missions; markers denote departure (green) and arrival (red)
    sites.}
    \label{fig:trajs_global}
\end{figure}

All 200 selected trajectories are collision-free, and the corridor constraint is active rather than slack: six of the 20 missions ride the 50~m limit in cruise (Fig.~\ref{fig:alt_m04}). An a-posteriori nearest-facet audit shows the clearance is honored everywhere except a single node of M06 on two seeds, which dips to 26~m. This is a consequence of the nearest-facet assignment lagging one iteration. The dip is a re-projection artifact, not a fundamental infeasibility: warm-starting an additional SCP pass from the converged M06 trajectory, so that the facet assignment is recomputed from the final geometry, restores the full 50~m margin in two iterations at a cost of 12~g on both affected seeds. This confirmatory pass is not applied to the catalog figures reported here. Representative trajectories are shown in Figs.~\ref{fig:trajs_p2} and~\ref{fig:trajs_global}.

\subsection{Warm-Start Ablation}
\label{subsec:ablation}

\begin{figure}[t]
    \centering
    \includegraphics[width=0.45\textwidth]{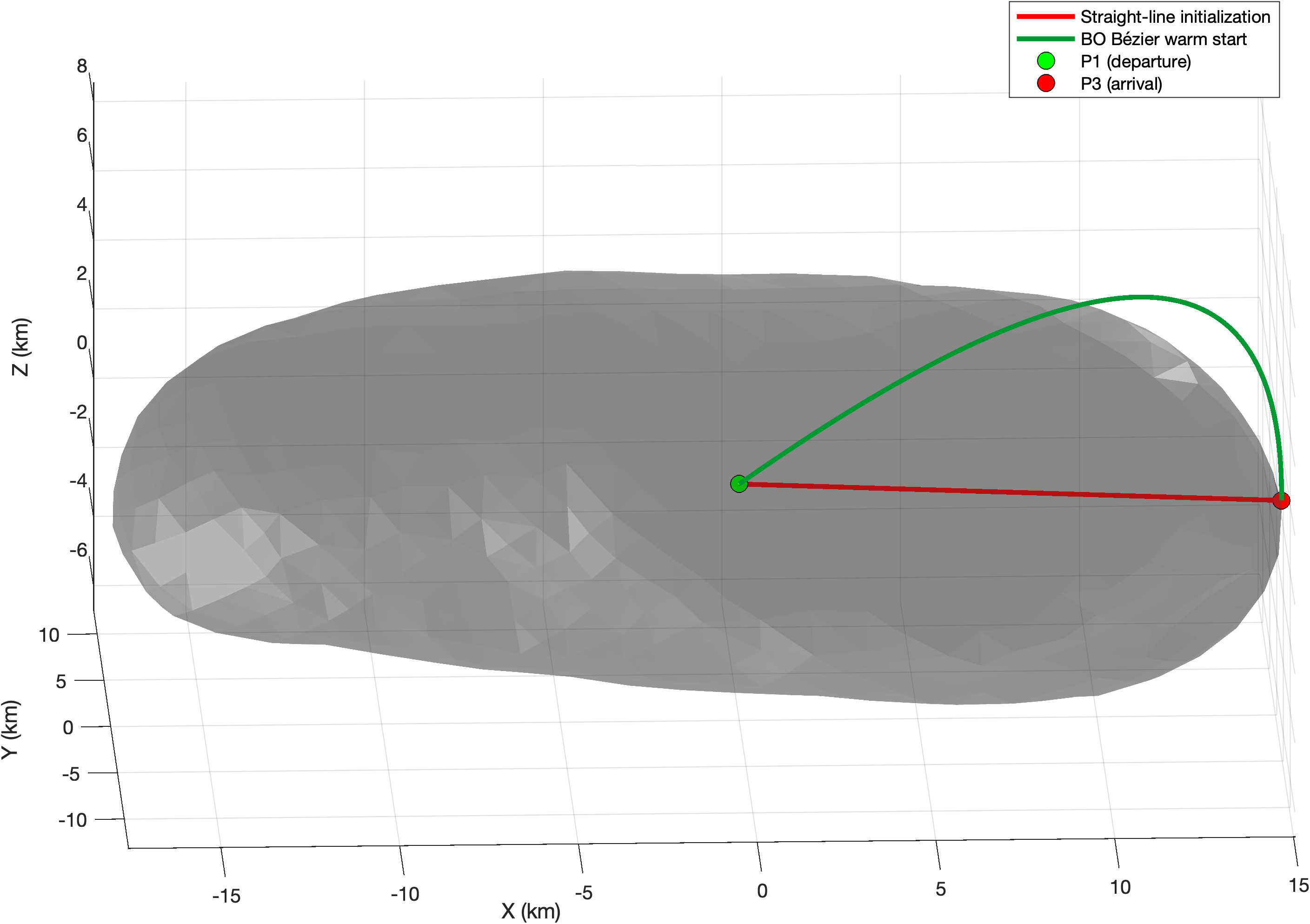}
    \caption{Straight-line versus physics-informed BO-optimized initialization for M02
    (P1$\to$P3): the straight-line reference (red) runs through the
    interior of Eros along its entire length, reaching $4.3$~km below the
    surface ($4.5$~km in the catalog's worst case), while the B\'ezier reference (green) clears the body.}
    \label{fig:ablation_vis}
\end{figure}

The physics-informed BO warm-start improves both reliability and convergence speed. In a seed-0 ablation re-solving every mission at its selected $t_f$ from a straight-line initialization, the straight line fails on one mission (M19, iteration cap) and averages 8.8 iterations versus physics-informed BO's 5.9 where both converge. The physics-informed BO reference is faster or equal on 17 of the 20 missions. Converged fuel costs are indistinguishable in the mean (1.53~kg for both), but the tails differ in both directions: straight-line lands in a poorer local basin on M15 ($+0.23$~kg) while finding a cheaper one on M04 ($-0.18$~kg). On a body this nonconvex, the initialization selects the basin; no simple reference dominates everywhere, but the physics-informed BO reference never fails to converge at the selected flight times. Figure~\ref{fig:ablation_vis} illustrates the underlying geometry: the straight-line reference starts deeply infeasible, whereas the B\'ezier reference starts clear of the surface.

\subsection{Seed Sensitivity}
\label{subsec:seed_dep}

\begin{figure}[t]
    \centering
    \includegraphics[width=0.45\textwidth]{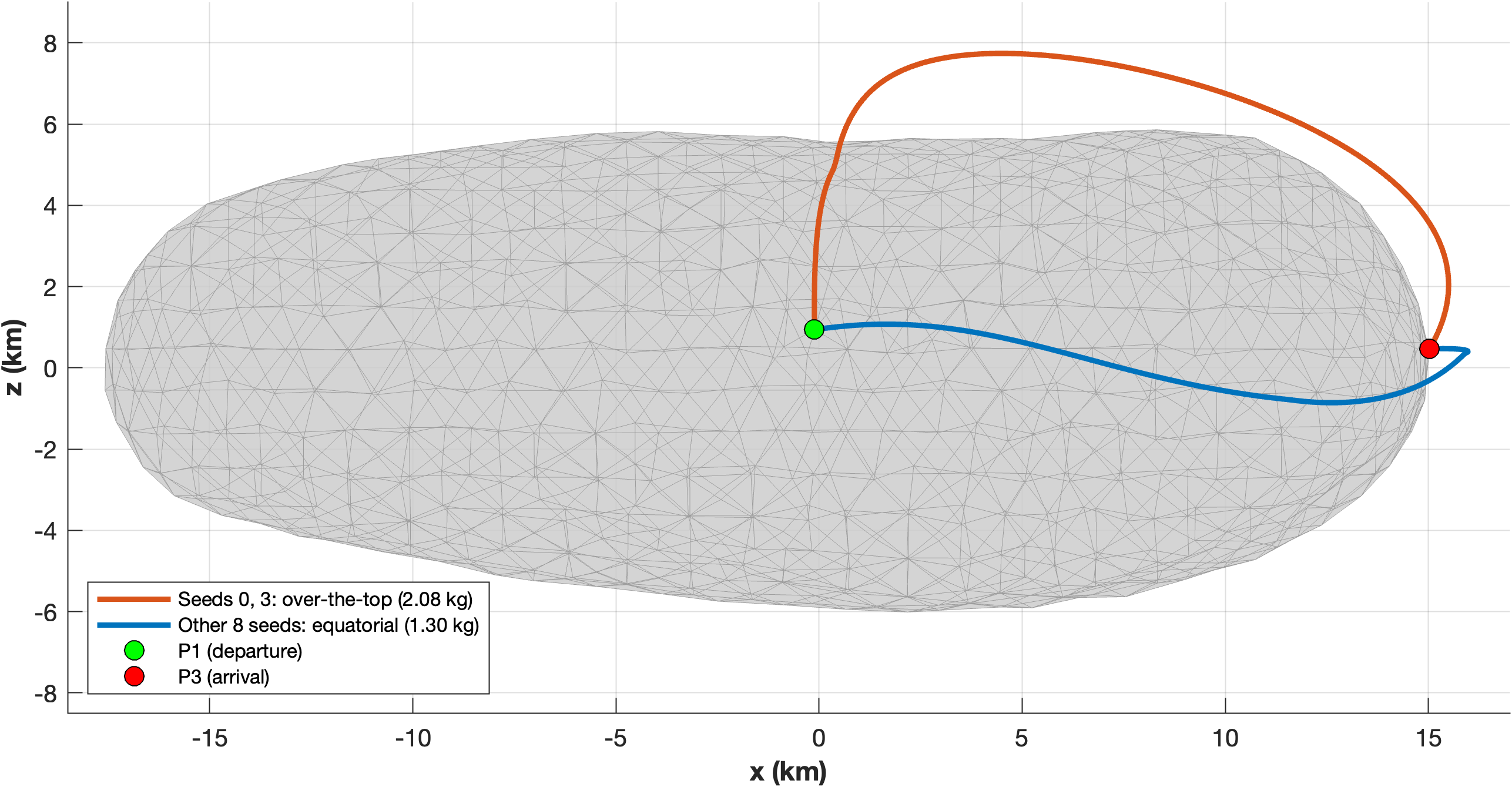}
    \caption{The 0.78~kg basin penalty is geometric: converged M02
    (P1$\to$P3) trajectories in $x$--$z$ projection. Seeds~0 and~3 climb
    over the body ($z \approx +7.6$~km at midpoint); the other eight
    seeds follow an equatorial corridor that wraps around the body in
    $y$ and only appears to cross it in this projection. The stochastic
    warm-start selects the basin; the local SCP stage refines within it.}
    \label{fig:basin}
\end{figure}

\begin{figure}[t]
    \centering
    \includegraphics[width=0.45\textwidth]{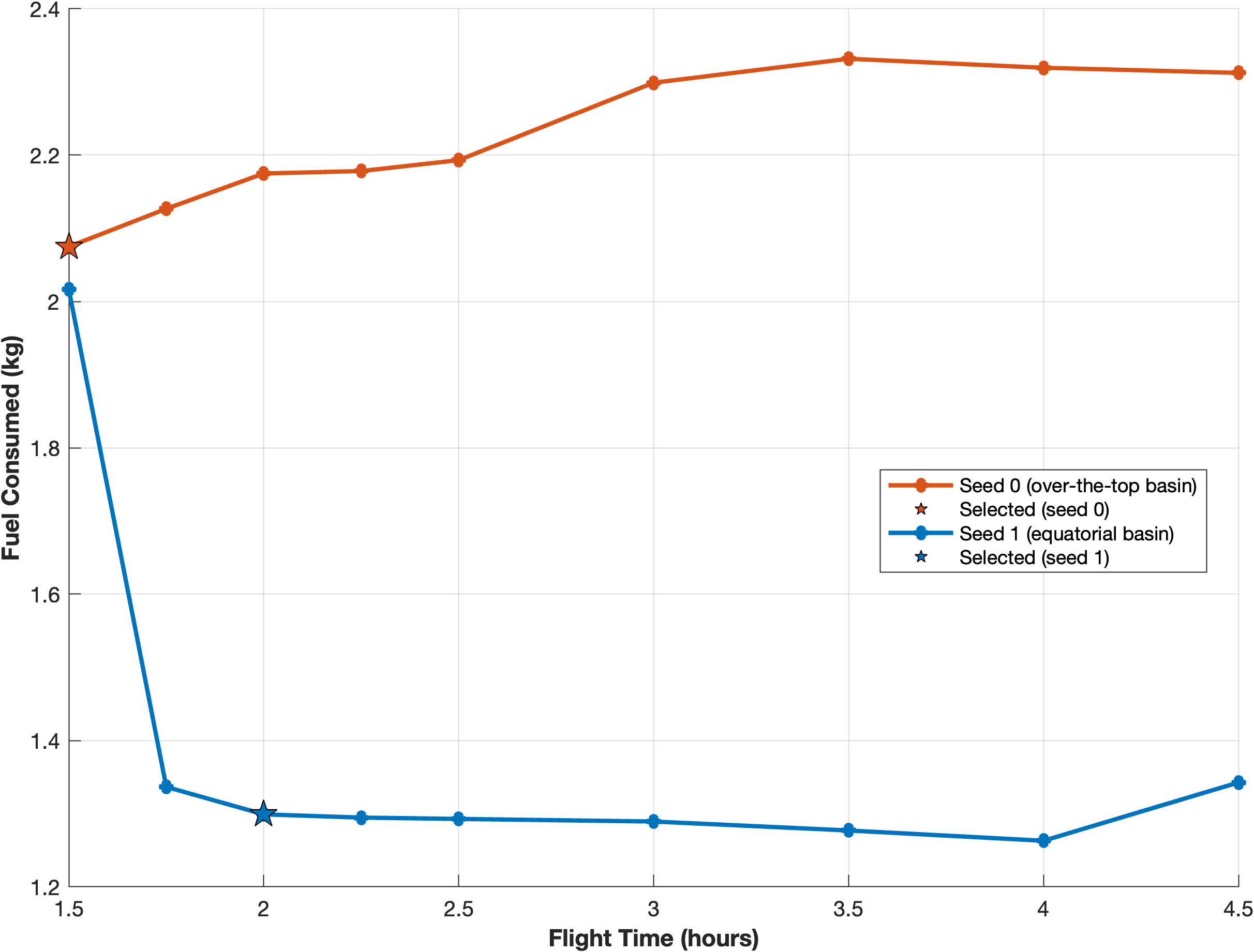}
    \caption{Basin capture shifts the entire Pareto front, not just the
    knee: time-of-flight sweeps for M02 (P1$\to$P3, 15.2~km) under seed~0,
    whose warm-start lands in the over-the-top corridor (2.08--2.33~kg at
    every $t_f$), and seed~1, which finds the equatorial corridor
    ($\approx 1.3$~kg beyond 1.75~h). Stars mark the utopia-distance
    selections.}
    \label{fig:pareto}
\end{figure}

Seed sensitivity, where present, acts on solution selection rather than solver convergence: 16 of the 20 missions are fully deterministic across all 10 seeds (identical $t_f$, fuel spread below 0.01~kg). The remaining four split into two distinct modes. The first is Pareto-level selection: M06 and M19 flip between grid-adjacent flight times (3.5 vs.\ 4.0~h and 4.0 vs.\ 4.5~h) whose utopia distances are nearly equal, so seed-driven variation in the underlying front determines which is selected, with fuel differences below 0.18~kg. The second is SCP-level basin selection, where the warm-start lands the solver in a different local basin of the nonconvex problem. On M02 (P1$\to$P3), seeds~0 and~3 converge to an over-the-top corridor (trajectory midpoint at $z \approx +7.6$~km) costing 2.08~kg, while the other eight seeds find an equatorial corridor at 1.30~kg. The two routes are geometrically distinct corridors around the body (Fig.~\ref{fig:basin}), and since each sweep point is warm-started from its neighbor, the entire Pareto front inherits the basin (Fig.~\ref{fig:pareto}). On M14 (P5$\to$P1), seed~2 mirrors the majority corridor about the equatorial plane (midpoint $z = -7.0$ vs.\ $+7.0$~km) at a modest $+0.11$~kg. Per-seed catalog totals span 29.68--30.58~kg ($1.1\%$ relative spread); excluding the two M02 basin-outlier seeds, the spread falls to $0.4\%$, indicating isolated basin capture rather than systematic bias.

This stochasticity is exploitable rather than merely tolerable: the seeds are independent, so a small ensemble of warm-starts constitutes a multi-start search over basins at ${\approx}10$~s per start, and selecting the cheapest converged solution eliminates the basin penalty---on M02, eight of ten seeds find the 1.30~kg equatorial corridor. Notably, the representative tour of Section~\ref{subsec:nominal} is composed entirely of deterministic missions and is seed-invariant.

\subsection{Robustness}
\label{sec:robustness}
\label{sec:mc}
\label{sec:perturb}

\textbf{Closed-loop Monte Carlo.} Closed-loop tracking is accurate to meters: over all $100{,}000$ runs the 99.7th-percentile terminal miss distance is 4.2~m (worst single run: 5.0~m), with per-mission mean misses of 0.10--3.4~m. Each nominal trajectory is tracked by a proportional--derivative (PD) guidance law ($K_p = 10^{-3}$~s$^{-2}$, $K_v = 5\times10^{-2}$~s$^{-1}$, $\zeta \approx 0.79$) under 500 stochastic simulations per trajectory with a 10~m ($1\sigma$) initial position dispersion and 2\% actuation noise ($\xi \sim \mathcal{N}(0,\,0.02^2)$, resampled at every integration step). At each guidance step:
\begin{equation}
    \mathbf{u}_{\mathrm{fb}} = K_p(\mathbf{r}_{\mathrm{ref}} - \mathbf{r}_{\mathrm{curr}}) + K_v(\mathbf{v}_{\mathrm{ref}} - \mathbf{v}_{\mathrm{curr}})
\end{equation}
\begin{equation}
    \mathbf{u}_{\mathrm{total}} = (\mathbf{u}_{\mathrm{nom}} + \mathbf{u}_{\mathrm{fb}})(1 + \xi)
\end{equation}
saturated to $T_{\mathrm{max}}/m_{\mathrm{nom}}$, where $m_{\mathrm{nom}}$ follows the nominal mass-depletion profile. State propagation uses fourth-order Runge--Kutta (RK4) integration at $\Delta t_{\mathrm{mc}} = \Delta t/5$ in the full nonlinear gravity environment. No success threshold is imposed; the miss-distance statistics themselves are the result: meter-level terminal accuracy under a simple PD tracker indicates that the SCP solutions are dynamically consistent and remain robustly trackable even where the clearance constraint is active. A nearest-facet audit of all $100{,}000$ closed-loop runs (exact point-triangle distance for any candidate under 150~m) confirms this directly: every mission/seed pair clears the surface by at least $49.6$~m except M06, where the facet-assignment lag of Section~\ref{subsec:nominal} persists into closed loop and the worst instantaneous clearance across all runs is $25.6$~m (seed~8, matching the $26$~m open-loop dip); terminal velocity error stays below $75$~mm/s (99.7th percentile $72$~mm/s) across the same runs, consistent with the zero-terminal-velocity constraint of Section~\ref{sec:problem}.

\textbf{Target perturbation.} The catalog solutions also respond smoothly to changes in the landing target. To show this, the arrival target is displaced by $\delta r = 15$~m along a uniformly random tangent-plane direction:
\begin{equation}
    \mathbf{r}_f' = \mathbf{r}_f + \delta r\,(\mathbf{b}_1\cos\theta + \mathbf{b}_2\sin\theta), \quad \theta \sim \mathcal{U}(0, 2\pi)
\end{equation}
where $\{\mathbf{b}_1, \mathbf{b}_2\}$ span the null space of $\hat{\mathbf{n}}_f^\top$. A mini-SCP (trust region tightened to 1~km) re-optimizes from the nominal Pareto solution across 10 perturbations per mission and seed (2000 re-optimizations). All converge. Averaged over missions, the maximum fuel deviation is 8.6~g, with a single-mission worst case of 44.6~g. Every perturbed problem returns the same corridor at essentially the same cost: no perturbation produces a basin jump, and the fuel cost varies smoothly with the target location.

\section{Conclusion and Future Work}
\label{sec:conclusion}

This paper presented a three-component framework for fuel-optimal collision-free surface-hopping trajectory design on asteroid 433~Eros: a physics-informed BO warm-start supplies SCP with a near-feasible B\'{e}zier reference, cutting the mean iteration count by a third and removing the convergence failure that straight-line initialization incurs, and a utopia-point time-of-flight sweep automates the time--fuel trade-off without introducing bilinear terms.

The presented 200-run campaign supports three conclusions: hopping mobility on Eros is propellant-affordable, a representative five-site tour consuming about a fifth of the propellant load; the randomness of the warm-start is an asset rather than a liability, with repeated seeds serving as a free multi-start search that maps the distinct solution basins of the nonconvex hop problem; and the optimized trajectories hold up in flight, flown to meter-level accuracy in closed loop by a simple PD law. Attitude dynamics, plume-surface interaction, and terrain uncertainty below the shape model resolution remain outside the study's scope.

Future work will apply the framework to hopping traverses of permanently shadowed lunar terrain and will compose catalog entries into complete multi-site tours through an optimal hop-sequence selector.

\section*{Acknowledgment}

The authors thank Dr.\ Cuma Yar\i m for his guidance and continued support throughout this work. The authors acknowledge using Claude to refine the grammar and enhance the expression of English.

\bibliography{bib}

\end{document}